# Coarse-grained approach to amorphous and anisotropic materials in Kinetic Monte Carlo thin-films growth simulations: a case study of TiO$_2$ and ZnO by Plasma Enhanced Chemical Vapor Deposition.

Jorge Budagosky,* Xabier García-Casas, Juan R. Sánchez-Valencia, Ángel Barranco, and Ana Borrás*

Nanotechnology on Surfaces and Plasma Group. Materials Science Institute of Seville (CSIC-US). C/ Vespucio 49, Seville 41092, Spain

Corresponding authors: jorge.budagosky@icmse.csic.es; anaisabel.borras@icmse.csic.es

## Abstract

The growth of amorphous TiO$_2$ and anisotropic-polycrystalline ZnO thin-films is studied by means of coarse-grained three-dimensional kinetic Montecarlo simulations under conditions typically encountered in Plasma Enhanced Chemical Vapor Deposition experiments. The approach developed considers fluctuations in the activation energy for surface diffusion of the coarse particles that are calculated on-the-fly and depends on the mesoscale local morphological/structural landscape. The basis of this approach --which is known to work well to simulate the growth of amorphous materials using a much simpler cubic grid-- has been extended in this work to reproduce not only the morphological characteristics and scaling properties of amorphous TiO$_2$ but also the growth of polycristalline ZnO with a good approximation, including the evolution of the film texture and textured-grain competition during growth and its dependence on experimental conditions. The results of the simulations have been compared with available experimental data obtained by X-Ray Diffraction, analysis of the texture coefficients, Atomic Force Microscopy and Scanning Electron Microscopy.

**Keywords:** ZnO, TiO$_2$, PECVD, Kinetic Monte Carlo, polycrystalline, texturization

## I. INTRODUCTION

Growth simulations tools are extremely important to reveal the mechanisms responsible for the structural features in thin films and low dimensional materials and the relationships with their physicochemical and functional properties. Thus, there is an ever-increasing interest in the development of simulation models with adaptability and flexibility to encompass predictive analysis on the deposition of simple and heterostructured materials. The major aim is to provide straightforward paths in the experimental routes towards the optimization of the properties and final performance in the targeted applications. Depending on the scale and nature of the features to be studied, the simulations usually range from those based on molecular dynamics (MD) [Chakraborty2013, Salvalaglio2015], kinetic Monte Carlo (KMC) [Burghaus2001, Weckman2018] or hybrid methods [Piana2006, Blackwell2013, Lloyd2018, Tonneau2021] to continuum methods [Smereka2005, Kelkar2014, Aebersold2019, Ramirez2020]. There is a wide range of technological applications for which the growth of thin films in multilayer systems (or even in more complex nanostructures) is a fundamental part. In many cases, the systems studied cover scales that are too large to be efficiently simulated by microscopic methods (e.g., MD and atomistic KMC), while at the same time show morphological characteristics that are difficult to simulate using continuous methods. It is in this *no-man's-land* where a coarse-grained (CG) treatment of the KMC method is often useful [Smith1996, Karabacak2001, Elsholz2004, Yanguas2006, Elsholz2007, Drews2004, Ralvarez2010a, Ralvarez2010b, Alvi2011, ABorras2012, Garnier2013, Moskovkin2014].

From a technological point of view, $TiO_2$ is part of a set of materials of intensive use in the manufacture of optoelectronic nano-devices, anti-reflective coatings, and antibacterial, biocompatible, anti-fogging, and self-cleanable surfaces. The control of the microstructure of the thin layers of this material is therefore an aspect that has been extensively studied before. Several simple models have been used previously to explain the anomalous roughness scaling behavior observed under various growth techniques [Karabacak2001, Elsholz2004, Yanguas2006, Elsholz2007, Ralvarez2010a, Ralvarez2010b, ABorras2012]. Since the growth methods usually employed in the manufacture of amorphous thin-films tend to operate in the low-temperature range and involve highly reactive surfaces, most of the models used to simulate the morphology of amorphous materials (e.g., $TiO_2$, $SiO_2$, $Nb_2O_5$, etc.) resort to various approaches that greatly restrict the range of growing conditions in which they can be applied. In this work, we have developed a KMC-based model that considers many of the general characteristics of those models but including fewer constraints. This permits it to cover a wider range of growing conditions, thus allowing its straightforward application to other materials. However, concerning the results shown here, these have focused on simulating growth processes under typical conditions found in the Plasma Enhanced Chemical Vapor Deposition (PECVD) technique. Such a deposition method has been applied during the last two decades in the development of functional coatings with applications in areas ranging from optics to biomaterials, including microelectronics and energy. Besides, during the last years, PECVD has been also extended to the fabrication of nanoscale materials, as nanoporous thin films and low dimensional nanostructures beyond graphene, CNTs, and nanowalls [Santhosh2018, Mao2015, McLeod2015, Ghosh2018], for instance, metal oxide nanorods and core@shell nanowires and nanotubes in combination with hard and soft template methods [Filippin2017a, Filippin2017b, Filippin2019, Filippin2020].

Thus, looking for the demonstration of the universal character of the proposed model, we have also approached the description of the crystalline anisotropic growth. Concretely, we have made emphasis in the simulation of the one-dimensional formation commonly observed in ZnO thin films grown by PECVD [RomeroGomez2010, Filippin2019]. The model is capable of reproducing, with a good approximation, the formation of polycrystalline layers of anisotropic materials, retaining the simplicity associated with the use of a cubic grid in the simulations. Also, we can tune the morphology and texture of the film throughout an anisotropy parameter that may be associated directly with microscopic surface phenomena and its dependence on the growth conditions. These results are of paramount interest in the application of PECVD methodology for the fabrication of polycrystalline layers in nanodevices, as third-generation solar cells, including Dye-Sensitized Solar Cells (DSSCs) and Perovskite Solar Cells (PSCs), photocatalytic electrodes, and in the development of piezoelectric and pyroelectric nanogenerators (PyNGs and PENGs [Boro2018, Hou2021, Lin2021, Pandey2021, Cao2021, Yang2021, Korkmaz2021]). In such applications, the precise control of the crystal growth parameters (crystal size, orientation, and texturization) allows for the design of nanomaterials with enhanced transport, optical, piezoelectric, and strain properties.

The paper is organized as follows, firstly, we describe the experimental setup used for the deposition and the analysis of the films, secondly, the theoretical background and the numerical method used in our simulations are described in detail. In the third place, we show and discuss the obtained predictions in comparison with the experimental results, and finally, we summarize the conclusions of our work.

## II. EXPERIMENTAL DETAILS

$TiO_2$ and ZnO thin films were fabricated by Plasma Enhanced Chemical Vapor Deposition (PECVD) at room temperature. Titanium tetraisopropoxide ($C_{12}H_{28}O_4Ti$) and Diethyl-Zinc (($CH_2CH_3)_2Zn$) precursors were purchased from Merck and used as delivered. Concretely, Ti precursor was inserted in the chamber by bubbling oxygen through a mass flow controller meanwhile the Zn precursor was inserted through a regulable valve. The base pressure of the chamber was lower than $10^{-5}$ and $10^{-4}$ mbar for the $TiO_2$ and ZnO, correspondently. The total pressure during the deposition was around $10^{-2}$ mbar in both cases achieved by supplying oxygen to the plasma reactor. The plasma was generated in a 2.45 GHz micro-wave Electron-Cyclotron Resonance (ECR) SLAN-II operating at 400 W for $TiO_2$ and 800 W for ZnO. SEM micrographs were acquired in a Hitachi S4800 working at 2 kV at working distances in the range of 2-4 mm. RMS roughness was characterized by Atomic Force Microscopy (AFM) in a Nanotech microscope in tapping mode. XRD patterns were recorded in Panalytical X'PERT PRO diffractometer model operating in the θ - 2θ configuration (Bragg-Brentano) and using the Cu Kα (1.5418 Å) radiation as an excitation source. The texture coefficients $T(hkl)$ were calculated applying the following equation:

$$T(hkl) = \frac{I(hkl)/I_0(hkl)}{\frac{1}{n}\sum I(hkl)/\sum I_0(hkl)} , (1)$$

Where $I(hkl)$ and $I_0(hkl)$ are the peak intensities associated with the $(hkl)$ family plane obtained in Bragg-Brentano ($\theta - 2\theta$) configuration for the samples and the randomly oriented reference pattern respectively (in this particular case the JCPDS card № 36–1451, for wurtzite) and n is the number of possible reflections.

## III. THEORETICAL MODEL

The proposed model for the growth simulations is implemented within an event-based KMC scheme, developed by Bortz-Kalos-Lebowitz (BKL) [BKL1975]. The code used to generate the simulation results shown in this work was developed in our group and is available from the authors upon reasonable request.

Some features of the model take the foundations from previous related works [Karabacak2001, Yanguas2006, ABorras2012, RAlvarez2010a, RAlvarez2010b]. This is described as follows:

**i)** The system is defined by a simple 3D cubic lattice with size $N_x \times N_y \times N_z$ and periodic boundary conditions (PBC) along the directions of the growth plane $(x, y)$. The 3D lattice is characterized by an integer array $G_{i,j,k}$ ($i = 1, \ldots, N_x$; $j = 1, \ldots, N_y$; $k = 1, \ldots, N_z$) in which, lateral overhangs and vacancies are allowed (see Fig. 1). Here, $G_{i,j,k} = 0$ for empty sites, $G_{i,j,k} = 1$ for substrate particles and $G_{i,j,k} > 1$ for particles of the film, with specific values that are related with the features of the model depending on whether one simulates an amorphous or an anisotropic material, as we will see later. In our model, the substrate particles are considered as immobile. Throughout this work, the substrate is set as a single flat layer of particles, i.e., $G_{i,j,k=1} = 1$ and $G_{i,j,k>1} = 0$. Each particle can perform a single site jump up to any available site in one of six possible directions associated to a cubic grid (left, right, back, front, up, down), except when a step is encountered in any direction of the 3D space. In that case, an extra jump to lower, upper or lateral sites is possible (see particles with label 3 in Fig. 1). The general constraint is that a particle must be bonded always to at least with one neighbor.

To keep things simple, the size of all types of coarse particles is the same and equal to the step size of the 3D lattice, $\Delta x = \Delta y = \Delta z = a_0$. In the case of the anisotropic material, each deposited coarse particle individually contains all the information regarding its crystallographic orientation with respect to the flat substrate. This feature will be elaborated on in more detail later. The size of the particles should be chosen as a compromise between computational performance and the pursued morphological detail [Drews2004].

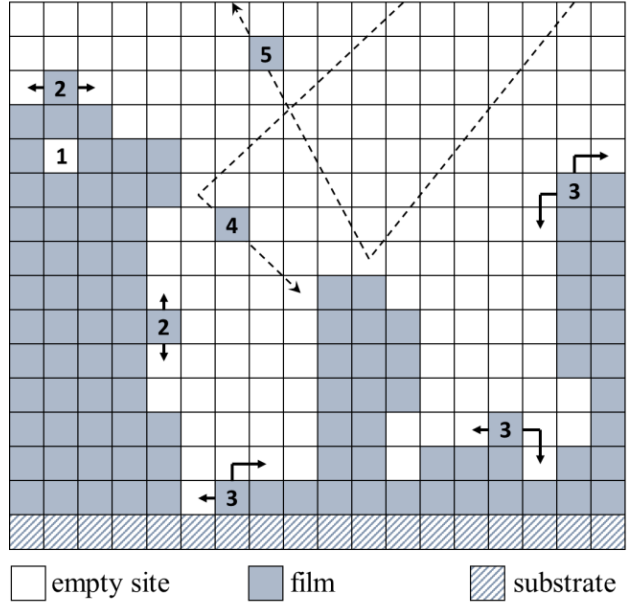

| empty site | film | substrate |

**Figure 1.** General scheme of the processes involved in the thin-film growth model developed in this work (For simplicity our 3D system is represented here in 2D). Several processes and configurations are included in the model: vacancies and overhangs are permitted (1; 2). A particle at the surface can jump randomly to any available site of six possible directions (2), except when a step is encountered (3), in that case, an extra jump is permitted. The deposition flux follows a distribution of trajectories for the incoming particles that may depend on the growth conditions. Once the particle reaches the surface, there is a possibility that it bounces and continues its journey with another trajectory until it reaches a new site (4) or until it reaches the roof of the simulation domain (5). If the latter occurs, that particle disappears and another particle is generated from scratch with a different trajectory and initial position at the roof.

ii) The deposition of new particles onto the surface is performed at a rate of $r_0 = F \times N_x \times N_y$, being $F$ the net deposition flux (in monolayers per second, ML/s, with a monolayer being defined as a single full layer of particles). In order to simulate a flux of particles that mimics typical conditions encountered in PECVD, the following features are considered: the incoming particles are generated at an initial position in the 3D mesh that is defined as the *roof* of our simulation domain ($i_0, j_0, k_0 = N_{roof}$), where $N_{roof} \leq N_z$ while the coordinates $i_0$ and $j_0$ are picked randomly from a uniform distribution. The term $N_{roof}$ is set as the maximum height of the surface profile plus one lattice unit. This choice is justified since we are trying to simulate thin-film growth at low chamber pressure. Under these conditions, the mean free path of the incoming particles in the gas phase is very large compared with the surface features of the film and one can saves computation time by avoiding the explicit simulation of the particle trip far above the surface since larger values of $N_{roof}$ are statistically equivalent [Messe2018].

The trajectory followed by the particle toward the surface is characterized by azimuthal $\varphi_{dep}$ and polar $\theta_{dep}$ angles (see Figure 2). These angles are obtained from an angular distribution that follows from a Maxwell-type distribution for the particle energies

[Yanguas2006]. Within this model, the azimuthal angle is sampled from a uniform distribution $\varphi_{dep} \in [0, 2\pi)$ while the polar angle is sampled via a distribution $g(\theta_{dep}, v_n)$ given by,

$$g(\theta_{dep}, v_n) = \frac{\int_0^{\theta_{dep}} F(\theta', v_n) d\theta'}{\int_0^{\pi/2} F(\theta', v_n) d\theta'} = \int_0^{\theta_{dep}} F(\theta', v_n) d\theta' \ , (2)$$

where

$$F(\Theta, v_n) = 1 - \cos\Theta \ \frac{exp(v_n^2 \cos^2\Theta)[1 + erf(v_n \cos\Theta)]}{exp(v_n^2)[1 + erf(v_n)]} \ . (3)$$

The specific shape of the distribution is controlled through a normalized average velocity, $v_n$, directed toward the surface. Small values of $v_n$ means that the angular distribution function spreads the incoming direction of the particles. On the contrary, larger values correspond to a more collimated flux directed toward the surface, as can be seen in Figure 2. Once the particle reaches the surface, there is a possibility (weighted by a sticking coefficient $S_0 \in [0,1]$ [Karabacak2001, Karabacak2011]) that it bounces and, after a specular reflection, continues its journey with another trajectory until it reaches a new site or until it reaches the roof of the simulation domain. In that case, the particle disappears and another particle with a different trajectory is generated.

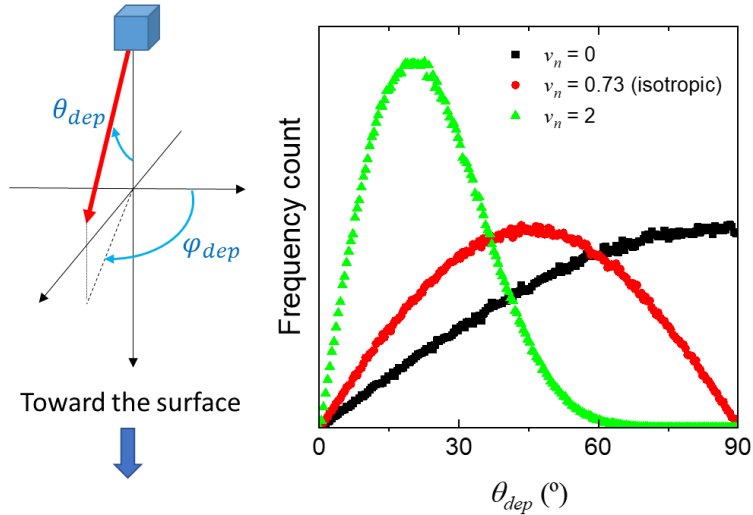

**Figure 2.** Angular distribution for the trajectories of the incoming particles toward the surface, considering three values of the normalized average velocity parameter $v_n$ (see text for details).

**iii)** In addition to deposition, our model includes surface diffusion: once at the surface, the movement of a coarse particle is simulated by jumping from lattice site to lattice site --as explained before-- via a random walk mechanism, characterized by a jump rate,

$$\Gamma_l = \frac{6D_0}{a_0^2} e^{-(\Delta E_l/K_B T)}, (4)$$

where the sub-index of $\Gamma_l$ labels a particular diffusion event from a list ($l = 1, \ldots, L$), $T$ is the substrate temperature, $K_B$ is the Boltzmann constant and the prefactor $6D_0/a_0^2$ define the time scale of the jump, being $D_0$ a parameter of the model interpreted as a microscopically averaged diffusion coefficient. The activation energy, $\Delta E_l$, for a diffusion event is calculated as,

$$\Delta E_l = nE_b + mE_w + \delta E = \Delta E_l^0 + \delta E, \quad (5)$$

where $E_b$ is the bond strength between particles of the film and $E_w$ between the film and the substrate, $n$ and $m$ are the number of film and substrate neighbors, respectively --Note that the index $l$ comprises the terms $n$ and $m$ in a single label, $l \equiv (n, m)$-- and $\delta E$ is an extra energy term that accounts for the structural features associated with the simulated material. This is the approximation used in [Elsholz2004, Elsholz2007, ABorras2012, RAlvarez2010], where the term $\delta E$ accounts for the energy fluctuations associated with the local structural fluctuations in amorphous lattices.

In its basic form (i.e., without the term $\delta E$ in the activation energy), the system has a very simple rate structure that depends on all possible $n + m$ combinations. However, here we assume that particles with five and six neighbors are immobile ($\max(n + m) < 5$). With this constraint, the size of the list of diffusion rates is set at $L = 14$. Below we will describe the calculation details of this term which are very different depending on if we are treating the amorphous or the anisotropic case.

The simulation begins with an initially bare substrate on which we begin to deposit particles following the procedure described above. On the other hand, the particles already deposited can move across the surface via surface diffusion. Once the particle reaches the surface, a $\delta E$ value is assigned to it based on the type of material and the particle neighborhood.

A. Amorphous material:

For the case of amorphous materials, $\delta E$ can randomly fluctuate spatially within an interval defined by the parameter $\Delta$ ($\delta E \in [0, \Delta]$). In our model, this energy interval is discretized in an array with $N_a$ elements evenly distributed, $\delta E = (\delta E_1, \delta E_2, \ldots, \delta E_\mu, \ldots, \delta E_{N_a})$. The index associated with this discretization helps us to add an identification label to the particle. In this way, a particle located at the site $(i, j, k)$ and having an associated value $\delta E_\mu$ ($\mu = 1, \ldots, N_a$) is labeled as $G_{i,j,k} = \mu + 1$. If the arriving particle is on the bare substrate and has no film neighbors (only substrate), $\delta E_\mu$ is set randomly from a uniform distribution. On the contrary, if one or more film neighbors are present $\delta E_\mu$ is obtained by averaging the $\delta E_\beta$ values of those neighbors. For diffusion events one follows a similar criterion: once the particle *jumps* to another position, the neighborhood of the arrival site is scanned to set the new $\delta E_\alpha$. Note that this procedure introduces implicitly lateral spatial correlations among the particles. The search radius for the averaging procedure may be changed to control the degree of correlation.

B. Anisotropic material (wurtzite):

As in the amorphous case, we simulate the surface diffusion in wurtzite-type materials like ZnO by means of an extra term $\delta E$ in the activation energy. However, in this case, $\delta E$ is defined as a function of the local morphology and crystallographic orientation. Again, a deposited particle with no film neighbors is labeled with a randomly chosen orientation of the crystal reference system ($x' = [11\bar{2}0], y' = [10\bar{1}0], z' = [0001]$), parameterized by a set of Euler's angles ($\theta, \varphi, \gamma$) defined with respect to the global reference system ($x, y, z$, where the $z$-axis is normal to the flat substrate, see Figure 3 for more details). If there are one or more film neighbors, we scan them and assign to the particle the crystallographic orientation of the majority of neighbors. In those situations where we have a tie between two or more options (e.g., if we have only two neighbors each with different Euler parameters), we randomly choose one of these. A similar criterion is applied for diffusion.

The process of assigning the crystallographic orientation to a coarse particle deposited on the bare substrate begins by discretizing the angles ($\theta, \varphi, \gamma$): $\theta = (\theta_1, \ldots, \theta_s, \ldots, \theta_{N_\theta})$, $\varphi = (\varphi_1, \ldots, \varphi_p, \ldots, \varphi_{N_\varphi})$ and $\gamma = (\gamma_1, \ldots, \gamma_q, \ldots, \gamma_{N_\gamma})$. Thus, one get a set of three numbers, $(s, p, q)$, that characterize a particular orientation of the crystal axes. This set is comprised of a single label, $\mu$, which help us to identify the particle by its orientation with respect to the substrate ($G_{i,j,k} = \mu + 1$).

As mentioned above, the calculation of $\delta E$ requires considering the local landscape around the particle that we want to move. The upper limit for this term is set also by $\Delta$. However, here this value corresponds to the extra energy added to the activation energy for the case of a particle deep inside the volume of the film. As can be seen in Figure 3, eight grid points around the particle contribute to $\Delta$: two of them aligned parallel to the local c-axis ([0001]) of the wurtzite and the remaining six forming an (approximate) hexagonal arrangement in the associated c-plane. We assume that the contribution of these *mesoscopic bonds* along the c-axis ($\delta E_\parallel$) may be different from that of the c-plane ($\delta E_\perp$). This difference is quantified using an *anisotropy ratio* $A_r$ ($\delta E_\parallel = A_r \delta E_\perp$):

$$\Delta = 2\delta E_\parallel + 6\delta E_\perp = 2(A_r + 3)\delta E_\perp. \quad (6)$$

The set of angles, contained in the label $\mu$, permit us to identify the location of the eight sites that must be checked to get $\delta E_\mu$. Obviously, empty sites ($G_{i,j,k} = 0$) are not included in the calculation:

$$\delta E_\mu = n_\parallel \delta E_\parallel + n_\perp \delta E_\perp, \quad (7)$$

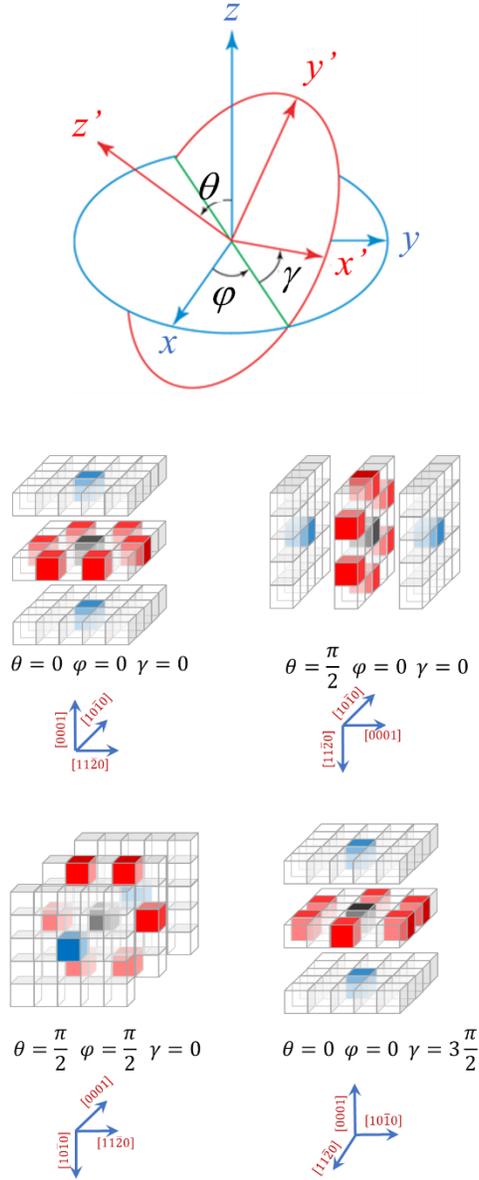

**Figure 3.** Example of four different localization of the neighbors considered in the calculation of $\delta E_\mu$, depending on the crystallographic orientation of the central particle (black cube). Each bond with the blue particles has strength $\delta E_\parallel = A_r \delta E_\perp$, whereas each one of the six bonds with the red particles has strength $\delta E_\perp$. Each one of these bonds is included in the calculation only if the site is occupied.

where $n_\parallel \leq 2$ and $n_\perp \leq 6$. Note in Fig. 3 that only the third and fourth nearest neighbors are considered (in fact, this separation may be tuned to even larger values). The reason for this is the angular resolution needed to reproduce approximately the hexagonal shape of c-plane neighbors arrangement and to provide a minimum resolution to reproduce the different crystal orientations during the simulation. This is one of the main limitations of the use of a cubic grid for our KMC simulator. Finally, the interaction between particles belonging to different textured-domains (grain boundaries) is addressed as follows: we assume that the strength of each mesoscopic bond during the evaluation of $\delta E_\mu$ is reduced by a factor that depends on the difference between the crystal orientation of the central $\mu$-particle with that of each of those far neighbors. Thus, equation (7) is restated as:

$$\delta E_\mu = \sum_{\rho=1}^{2} \delta E_\parallel M_{\mu,\rho} + \sum_{\sigma=1}^{6} \delta E_\perp M_{\mu,\sigma}, \quad (8)$$

with $M_{\mu,v} = \cos^2(\Delta\theta_{\mu,v}) \cos^2(\Delta\varphi_{\mu,v}) \cos^2(\Delta\gamma_{\mu,v})$, being $\Delta\theta_{\mu,v}$, $\Delta\varphi_{\mu,v}$ and $\Delta\gamma_{\mu,v}$ the difference between the Euler's angles of the crystal orientation of particles $\mu$ and $v$.

During the simulation, the selection of a particular diffusion event is performed through the standard BKL scheme excluding the term $\delta E_\mu$. This is done in order to avoid increasing the size and complexity of the events list of the algorithm. Once an $l$-event with rate $\Gamma_l$ is selected, the term $\delta E_\mu$ is included by means of an acceptance-rejection method [Saum2009]. The procedure is the following: since $\delta E_\mu \in [0, \Delta]$, the addition of this term to the energy barrier means that one get a new rate, $\widetilde{\Gamma}_l$, that its lower than the original ($\widetilde{\Gamma}_l \leq \Gamma_l$). In that sense, the rate $\Gamma_l$ represent the upper limit in an interval of possible values per each l-type event. Once we get $\widetilde{\Gamma}_l$ a uniformly distributed random number $\zeta_1 \in [0,1)$ is generated and compared with the ratio $\widetilde{\Gamma}_l/\Gamma_l$. If $\zeta_1 < \widetilde{\Gamma}_l/\Gamma_l$ the jump is performed. On the contrary, another particle is selected for jumping following the same procedure. The computation time wasted in this acceptance-rejection procedure will depend on how small is the ratio $\widetilde{\Gamma}_l/\Gamma_l$.

Once a particle is selected for jumping, the jump direction is chosen by searching the available neighbor sites. In general, the arrival site is selected randomly except when a step is encountered. In that case, the probability of the jump to that site is reduced by a factor $e^{-(\zeta_2/K_B T)}$, being $\zeta_2$ a uniform random number ($\zeta_2 \in [0, E_s)$, where $E_s$ is an energy parameter of the model). Note that, although Schwöebel-like barriers are found to be negligible in amorphous materials [Yang1996], we consider it necessary to include a step-barrier here to hind the intrinsic cubic nature of our coarse-grained computational array under a wider range of growth conditions, where diffusion processes play a major role. Nevertheless, the inclusion of randomness for this barrier permits us to approach the lack of slope selection characteristic of amorphous film morphologies.

Finally, to visualize the simulations, the coordinates of all the particles are saved in *xyz* format. Thus, the files were loaded in Ovito [Ovito2010], a 3D visualization software suitable for this purpose.

## IV. RESULTS AND DISCUSSION

A. Nanoporosity and scaling behavior of (amorphous) TiO$_2$ thin films:

In the case of TiO$_2$, in all the simulations we have considered an array of $201 \times 201 \times 1001$ with a grid step of $a_0 = 2.5$ nm, which gives us a surface area of size $500 \times 500$ nm$^2$. The nominal growth rate (nominal thickness versus time) has been set at 4 nm/min (0.027 ML/s). The rest of the parameters of the model have been set to $D_0 = 3 \times 10^3$ nm$^2$/s, $E_b = 0.08$ eV, $E_w = E_b$, $E_s = E_b/2$ and $\Delta = E_b/2$ (for the discretization of the fluctuating energy term we set $N_a = 41$). Finally, all the simulations are carried at room temperature (RT).

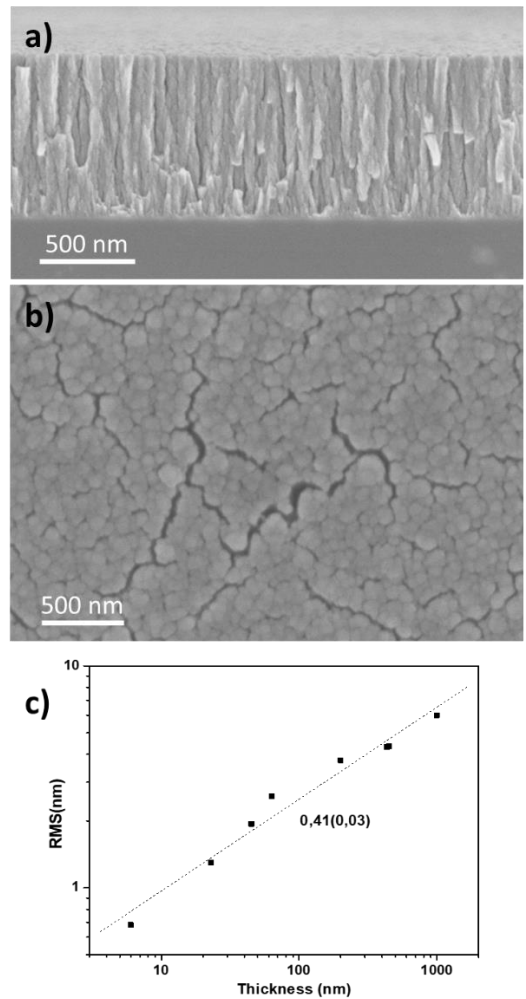

**Figure 4**. SEM cross-section (a) and top-view (b) micrographs showing the columnar-like morphology of the TiO$_2$ PECVD thin films deposited at room temperature. Experimental estimation of the ß parameter from the RMS values obtained by AFM characterization of films with increased thickness.

Figure 4 a-b) gathers the characteristic SEM cross-section and top view micrographs of a TiO$_2$ thin film fabricated by PECVD under the experimental conditions detailed in the Methods section, RMS values as a function of the thin film thickness are represented in panel c). In addition, Figure 5 a-d) shows the evolution of the morphology of the film as a function of the amount of material deposited (coverage), considering two different angular distributions for the flux of arriving particles. In both cases, we have set the value of the sticking coefficient $S_0$ at 1 and 0.65 [Karabacak2001]. We can verify that in all cases, the morphologies depict a columnar-like formation in good agreement with the SEM micrographs in Figure 4. On the other hand, when the flow of arriving particles is more dispersed ($v_n = 0$), the separation between these columns tends to be larger. Also, although in both cases the diameter of the columns tends to increase with the thickness of the film, in Fig. 4 (a) this increase is much more evident. This last result is known to be a consequence of the shadowing effect and is directly related to the angular dispersion of the particle flow [Yanguas2006, ABorras2007, ABorras2012].

In general, several of the morphological characteristics observed can be quantified in terms of the R.M.S roughness:

$$w(t) = \sqrt{\langle(h(r_\perp,t) - \bar{h}(r_\perp,t))^2\rangle_{r_\perp}}, \quad (9)$$

Where $\langle...\rangle_{r_\perp}$ denote average with respect to in-plane coordinates $r_\perp$, $h(r_\perp,t)$ is the local height at $r_\perp$ and $\bar{h}(r_\perp,t) = \langle h(r_\perp,t)\rangle_{r_\perp}$. Starting from $w(t)$, the dynamic properties of the film can be described according to a power law, $w(t) \sim t^\beta$ (or $w(h_0) \sim h_0^\beta$, being $h_0$ the coverage), characterized by an effective growth exponent $\beta$. The different growth regimes are quantified in this way in terms of the value of $\beta$ [Elsholz2004, Elsholz2007, ABorras2007]. Values of $\beta > 0.5$ obtained experimentally for the case of TiO$_2$, SiO$_2$ and organic films are attributable to the amorphous character of the material, although later it has been observed that the angular distribution of trajectories of the arriving atoms also plays a very important role via the shadowing effect [Yanguas2006, ABorras2007, ABorras2012] (see Fig. 4 (c)). This can be observed in Fig. 5 (c), where the effective growth exponents differ significantly from each other. In the case of the isotropic angular distribution ($v_n = 0.73$) it is even observed that the values obtained are around 0.5 for the growth conditions (growth rate and temperature) considered here.

Under these conditions, when we consider a sticking coefficient equal to 0.65, we obtain a value of $\beta$ similar to that obtained in Fig. 4 (c). Note that the increase in angular dispersion ($v_n = 0$) translates into a larger response of the morphology to changes in $S_0$. In the case of the RMS roughness, the difference obtained between the cases $S_0 = 0.65$ and $S_0 = 1$ for $v_n = 0$ at maximum thickness is approximately three times (~35 nm) that the one obtained when $v_n = 0.73$ (~10 nm). However, the situation with the effective growth exponent is different. In this case, larger $\beta$ values show smaller $S_0$-dependent changes for $v_n = 0$ than for $v_n = 0.73$. The foregoing shows that the proportionality constant in $w(h_0)$ is strongly dependent on the shadowing and plays an important role when one faces deposition flows that are more spreads away to the normal-to-surface direction. Thus, when shadowing tends to be dominant with respect to diffusion events, the morphology of the film tends to be more sensitive to small changes in the material-related properties. As was found in [ABorras2007, ABorras2012], no surface diffusion of TiO$_2$ *ad-species* is expected under the Full O$_2$ plasma growth conditions of Fig. 4 (a-c), so a high sticking coefficient and a low surface diffusivity yield the observed columnar growth when these conditions are combined with isotropic angular dispersion of the flow of incoming particles, typical of the PECVD technique. This conclusion is complemented by our results, which show that for an angular dispersion shifted toward more horizontal trajectories, the shadowing effect is boosted, and the columns grow faster, with a wider dispersion of heights and with more space between them.

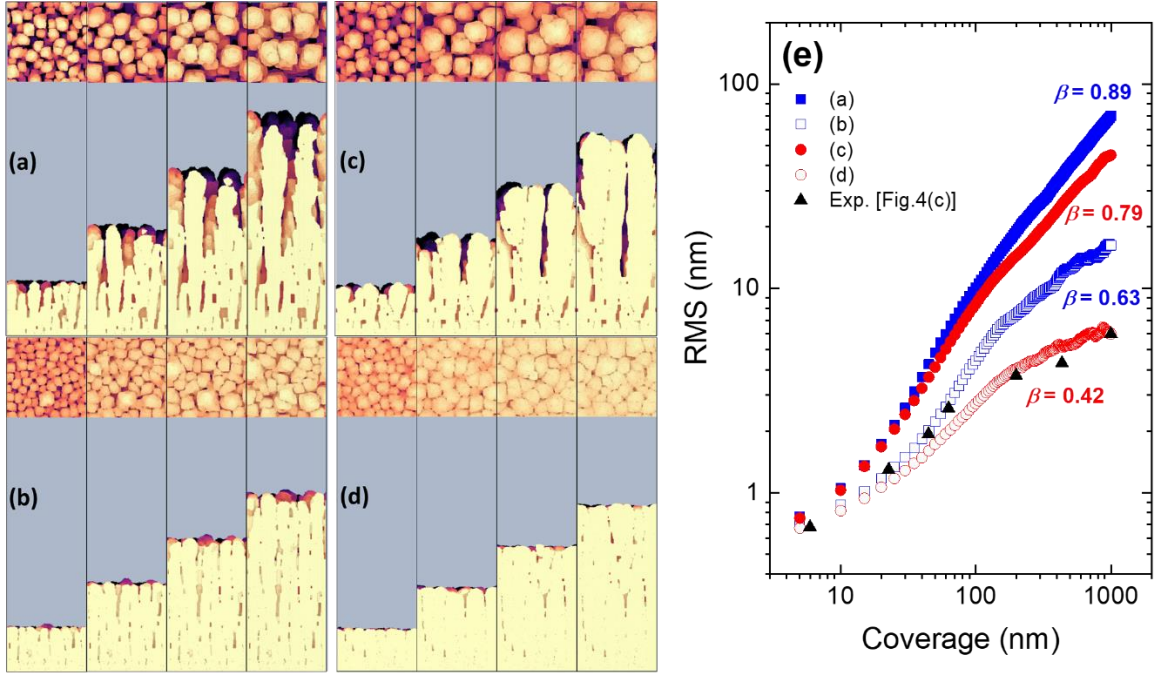

**Figure 5.** Simulated film morphologies at coverages 250 nm, 500 nm, 750 nm and 1000 nm for (a) $v_n = 0$ ; $S_0 = 1$, (b) $v_n = 0.73$ ; $S_0 = 1$, (c) $v_n = 0$ ; $S_0 = 0.65$ and (d) $v_n = 0.73$ ; $S_0 = 0.65$. Evolution of the calculated R.M.S. roughness as a function of thickness for cases (a-d). The black triangles are the experimental data of Fig. 4(c). The growth exponents obtained for each case are shown. The rest of the simulation parameters are described in the text.

*B. Microstructure and texture properties of polycrystalline ZnO thin films*

For the case of wurtzite-type ZnO, we have considered a grid with $201 \times 201 \times 1001$ elements, a grid step of $a_0 = 1.625$ nm and a surface of size $375 \times 375$ nm². The energy parameters have been set to $D_0 = 3 \times 10^2$ nm²/s, $E_b = 0.09$ eV, $E_w = E_b$, $E_s = E_b/2$ and $\Delta = 3E_b$. The values of the $v_n$ and $S_0$ parameters have been set to 0 and 1, respectively. Regarding the initial conditions, the code developed allows establishing a selection of permitted textures for the particles that arrive at the surface of the substrate. In this way, when a particle is deposited on the substrate, the crystallographic orientation that will be assigned to it will be chosen randomly within the selection of allowed textures. The objective of this control possibility is to be able to play with the possible influence of the substrate itself on the texture of the first clusters in the initial instants of growth. The origin of the texture shown by ZnO thin films under different growing conditions and the question as to whether this is defined early on or is progressively defined during the growth through a process of evolutionary competition of textured grains/domains is still under discussion. For this work, in the examples shown below we have considered it sufficient that the particles that arrive at the surface can be labeled with one of only three possible textures with equal probability: [002] ($\theta = 0, \varphi = \zeta_a 2\pi, \gamma = \zeta_b 2\pi$), [101] ($\theta = \pi/3, \varphi = \zeta_a 2\pi, \gamma = \pi/2$) and [100] ($\theta = \pi/2, \varphi = \zeta_a 2\pi, \gamma = \pi/2$), being $\zeta_{a,b}$ two uniformly distributed random numbers within the interval [0,1]. Finally, in order to study texture-related characteristics of the simulated films and to be able to make better comparisons with experiments, we calculate the texture coefficients in our samples. To obtain the texture coefficient for a particular crystal orientation, e.g. [002], we sum up all those particles labeled with that texture and divide the result by the total number of particles of the film (i.e., excluding the particles of the substrate). These values have been compared with the experimental Texturization coefficients obtained from the XRD patterns (see Methods). Figure 6 shows top-view and cross-section images of a ZnO film fabricated on substrates at room temperature conditions under oxygen plasma and growth rate of 10 nm/min (a-b), the corresponding XRD pattern acquired under Bragg-Brentano configuration normalized to the higher peak for different thicknesses (c) and the calculated texturization coefficients (d). Figure S1 in the Supporting Information Section gathers the XRD patterns for ZnO thin films deposited under different conditions of substrate temperature, growth rate, and thickness.

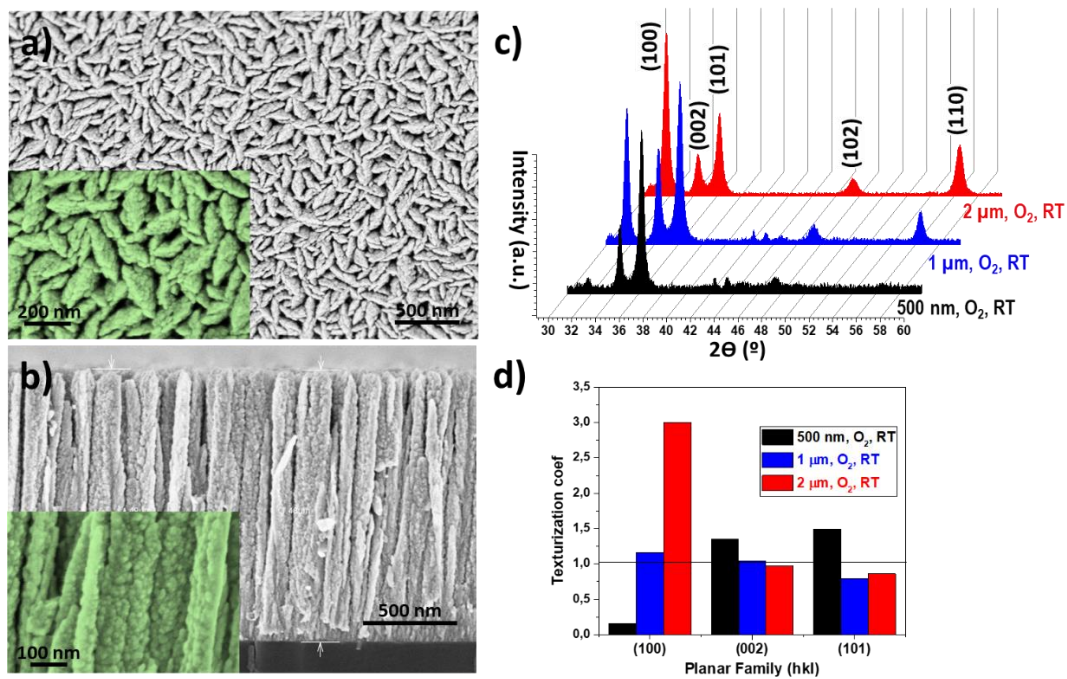

**Figure 6**. SEM top-view (a) and cross-section (b) micrographs showing the columnar-like morphology of the ZnO PECVD thin films deposited at room temperature under oxygen plasma for different thicknesses as labeled. (c) XRD patterns acquired under Bragg-Brentano configuration (indication on the main planes for ZnO wurtzite) (d) and texturization coefficient for the samples deposited on fused silica substrates.

Figure 7 shows the morphology obtained for different ZnO films considering several values of the anisotropy ratio, $A_r$. In these simulations, the temperature of the sample has been set at RT and the nominal growth rate at 5 nm/min [Fig. 7 (a,b)] and 10 nm/min [Figs. 7 (c,d)], which correspond to growth rates of approximately 10 nm/min and 20 nm/min, respectively, if we calculate them in terms of the film thickness (as in the experiments) rather than the nominal coverage. In Fig. 7(a,c) the morphology of each film is highlighted, while in (c,d) the chosen colors make it possible to highlight the textures of the different domains (grains) that appear in each case. The first feature that can be verified is that the values of $A_r$ that is much larger than 1 --which means a higher growth rate along the c-axis of the wurtzite than in a direction perpendicular to it-- give rise to a film that is almost totally [002]-textured. The columnar morphology associated with this texture is reasonably consistent with the results in Figure 6 and previous results for thin films fabricated under similar growth conditions [RomeroGomez2010] (see also Figure S1). Due to shadowing effects, the observed columns are narrow at the base and tend to increase in diameter with increasing thickness. As we reduce the value of $A_r$ for a growth rate of 5 nm/min (real growth rate ~10 nm/min), an increasingly abundant presence of [101]-textured domains is observed. The consequences in morphology are quite evident: the nano-columns tend to show a slightly tilted axis and their cross-section turns out to be anisotropic, showing a rather pronounced elongation perpendicular to the [002] axis. As we approach toward $A_r = 1$, i.e., when the growth rates parallel and perpendicular to the c axis are equal, the texture of the film changes again rapidly and becomes mostly [100]. In this region, the nano-columns continue to show an evident anisotropy in their cross-section. On the other hand, these recover the verticality previously observed when the texture was [002]. The elongated cross-sectional shape of the nanocolumns for $A_r \leq 1.5$ agrees very well with the morphology shown in Figure 6 (a). Likewise, their slight tilt, as observed in Figure 6 (b), coincides with the presence of [101]-textured domains in the film [see Fig. 6 (c-d)]. This result is consistent with what was observed in the simulations for values of $A_r$ between 1 and 1.25. Let look now the results when the growth rate is set at 10 nm/min (real growth rate ~20 nm/min). With respect to the previous examples, the effects of a larger growth rate are evident not only in the morphologies but also in the textures of the film. Particularly around $A_r$~1.5. This time the film shows a direct transition from mostly [100]-textured to mostly [002]-textured as we move from $A_r = 0.5$ to $A_r = 2.5$, without passing throughout any [101]-textured stage.

The previous results are summarized in Figure 8, where the texture composition of each film is plotted as a function of the coverage. At the smallest (and largest) values of $A_r$, the texture coefficients seem to be insensitive to changes in growth conditions. It is in the range $A_r = 1.0 - 2.0$ that growth rate becomes relevant for the final texture of the films. In this region, it is possible to observe the competitive behavior of textured domains in some cases.

Qualitatively, the behavior of the components [101] [100] of the texture coefficients obtained in our simulations is similar to the results obtained in the real samples under similar growth conditions [Fig. 6 (c-d)]. However, the evolution of the component [002] in the simulations is somewhat different from the experimental one. In our simulations, this component practically disappears at later stages of growth. This could be a consequence of the interaction with the substrate: just at the beginning of growth, the bare substrate surface would tend to favor the formation of nuclei or islands with texture [002] over textures [101] and [100]. Later, as growth advanced, evolutionary competition would favor the latter depending on the growth conditions ($A_r \leq 1.5$) but still with a noticeable presence of [002] domains. In principle, this may be addressed in our simulator by setting the arrival probability of [002]-textured particles higher than the other textures. This aspect will be addressed in greater detail in incoming work.

Finally, in Fig. 9, the texture coefficient at maximum coverage for all the cases in Fig. 8 is shown as a function of $A_r$. Here, it is more evident what was mentioned above: when the values of the anisotropy ratio are within the range $A_r = 1.0 - 2.0$, the final texture of the film is more sensitive to changes in growth conditions, like the growth rate. Comparing our results with the amplitudes of the peaks in the XRD patterns of ZnO films shown in Refs [VegaPoot2014] (Fig 2), Fig. 6 and Fig. S1, we found a reasonable agreement with the relative values of our texture coefficients in Fig. 9 for values of $A_r$ between 1 and 1.5. In addition, our results at $A_r \geq 2$ agree well with the results shown in [RomeroGomez2010]. It is interesting to note that in the former case, the plasma composition was 90% $O_2$ and 10% $H_2$, whereas in the latter the sample is grown with pure oxygen plasma and for different power of the microwave source. This gives us a hint about the associated values of $A_r$ on each case, suggesting that this parameter might be connected with specific microscopic surface features during the growth, that are related to the plasma composition.

**Conclusions**

In conclusion, in this work, we have developed a model based on the lattice KMC method to simulate the growth of thin films under conditions typical of vapor or vacuum deposition methods, such as PECVD. Using a CG approximation, the developed model allows the simulation of amorphous and anisotropic materials starting from a cubic three-dimensional array, which makes it particularly useful for reproducing morphological characteristics at scales beyond the current possibilities of more sophisticated models, like MD or atomistic KMC, using moderate computational resources. The growth of $TiO_2$ at room temperature has been precisely simulated gaining insight of the mechanism behind the formation of columnar-shape morphologies and a detailed view on the height dispersion and distance between columns. In the case of anisotropic materials, the model allows simulating the formation of grains with different textures within the same film (polycrystalline growth). We have chosen Zinc Oxide with a wurtzite structure as a proof-of-principle for this work. However, the approximation used for the hexagonal structure associated with wurtzite could be extended to other types of structures. At a qualitative level, we have found that the texture coefficients obtained for ZnO agree reasonably well with our experimental results and with some experimental results found in the literature, under growth conditions similar to those considered in this work. Regarding this, currently, a more detailed experimental study of the growth of ZnO under different growth conditions is in the process by our group and in which extensive use of the present simulator will be made.

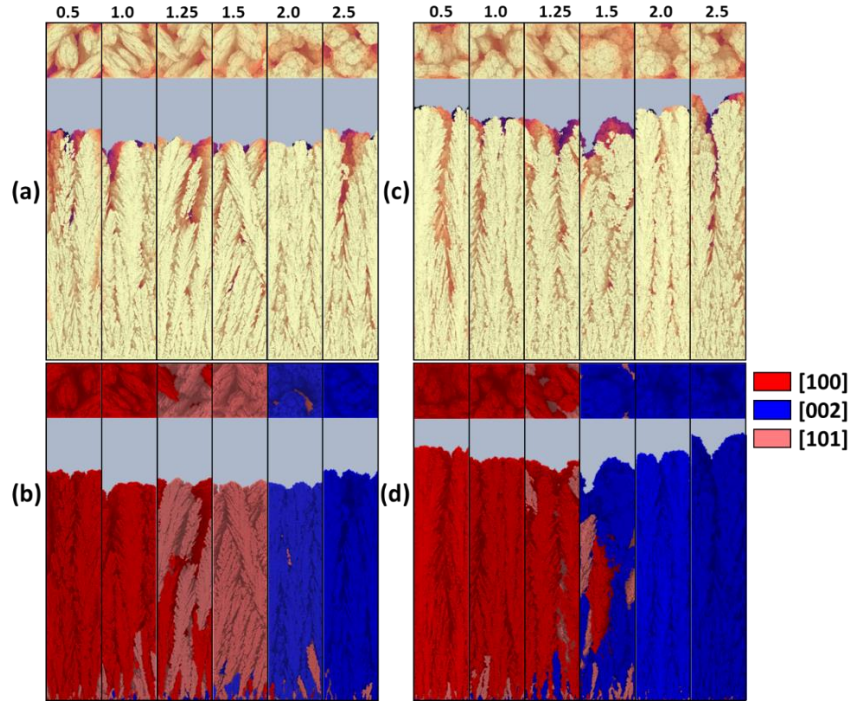

**Figure 7.** ZnO thin-film morphologies for several values of $A_r$ (shown at the top of each plot) for a nominal growth rate of 5 nm/min (a-b) and 10 nm/min (b-c), both at RT. The color selection for the plots in (a) and (c) is meant to highlight the morphology of the samples while in (b) and (d) the colors selection is used to identify visually the different textured grains present in a single film. The details about simulation parameters are described in the text.

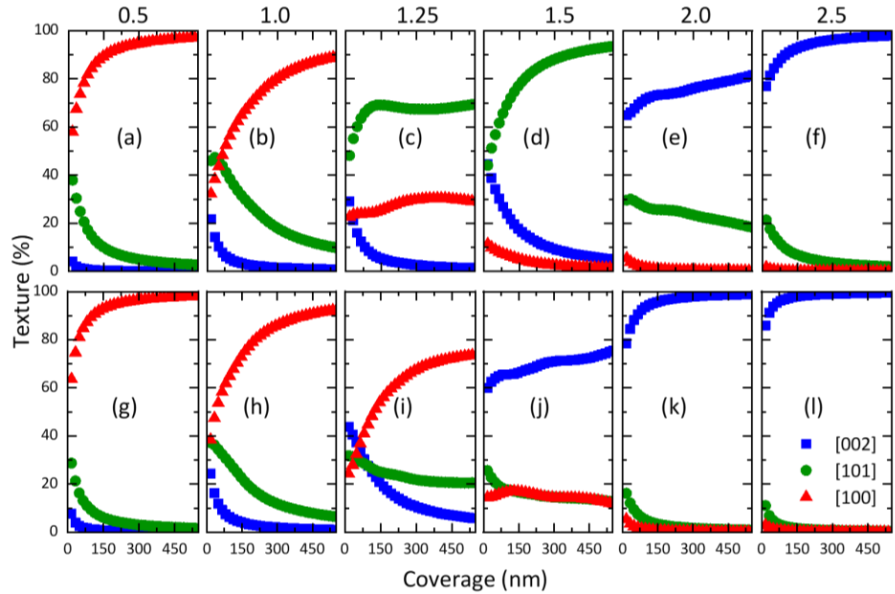

**Figure 8.** Evolution of the texture coefficients for the simulations shown in Fig. 7 as a function of (nominal) coverage, evaluated for six values of $A_r$ and considering the two different growth rates: (a-f) 5 nm/min and (g-l) 10 nm/min.

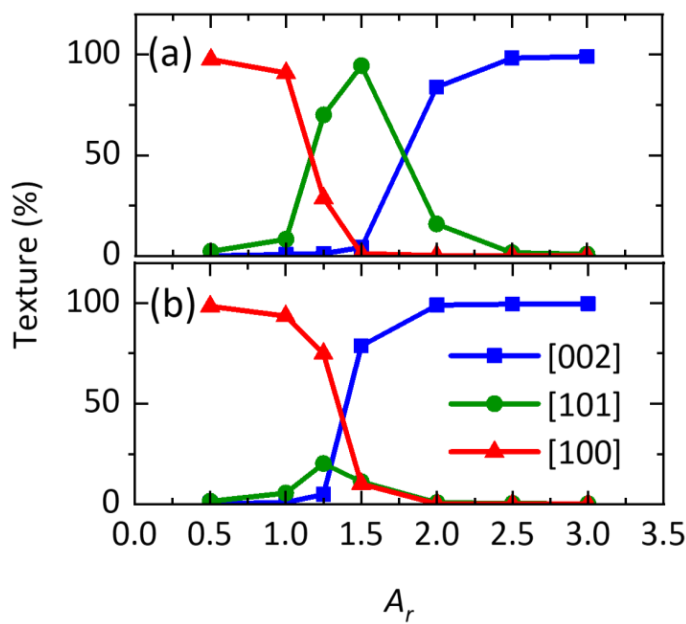

**Figure 9.** Texture coefficients at maximum thickness for the simulated ZnO thin films as a function of $A_r$ (T=RT) for two different growth rates: (a) 5 nm/min and (b) 10 nm/min.

**Acknowledgments**
The authors thank the AEI-MICINN (PID2019-110430GB-C21 and PID2019-109603RA-I0), the Consejería de Economía, Conocimiento, Empresas y Universidad de la Junta de Andalucía (PAIDI-2020 through projects US-1263142, ref. AT17-6079, P18-RT-3480), and the EU through cohesion fund and FEDER 2014–2020 programs for financial support. J.R.S.-V. thank the University of Seville through the VI PPIT-US and (J.R.S.-V.) the Ramon y Cajal Spanish National programs. The projects leading to this article received funding from the EU H2020 program under the grant agreements 851929 (ERC Starting Grant 3DScavengers) and 899352 (FETOPEN-01-2018-2019-2020—SOUNDofICE).

# SUPPORTING INFORMATION

**Coarse-grained approach to amorphous and anisotropic materials in Kinetic Monte Carlo thin-films growth simulations: a case study of $TiO_2$ and ZnO by Plasma Enhanced Chemical Vapor Deposition.**

Jorge Budagosky,* Xabier García-Casas, Juan R. Sánchez-Valencia, Ángel Barranco, and Ana Borrás*

Nanotechnology on Surfaces and Plasma Group. Materials Science Institute of Seville (CSIC-US). C/ Vespucio 49, Seville 41092, Spain

Corresponding authors: jorge.budagosky@icmse.csic.es; anaisabel.borras@icmse.csic.es

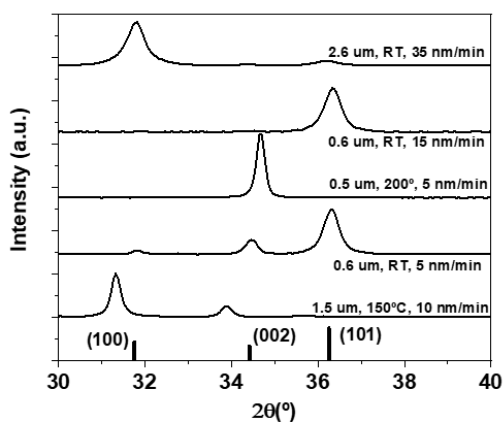

**Figure S1.** XRD patterns corresponding to different experimental conditions and thicknesses for ZnO fabrication by PECVD